\input{aipcheck}

\documentclass[
final            % use final for the camera ready runs
  ]
  {aipproc}

\layoutstyle{6x9}

\begin{document}

\title{A planet of an A-star: HD15082b}


\author{Eike W. Guenther}{address={Th\"uringer Landessternwarte Tautenburg, 07778 Tautenburg, Germany}}
\author{Andrew Collier Cameron}{address={SUPA, School of Physics and Astronomy, University of St. Andrews,
        North Haugh, St. Andrews, Five KY16 9SS, UK}}
\author{Barry Smalley}{address={Astrophysics Group, Keele University, Staffordshire ST5 5BG, UK}}
\author{Florian Rodler}{address={Instituto de Astrof\'\i sica de Canarias, C/V\'\i a L\'actea, s/n, E38205 -- La Laguna (Tenerife), Spain}}
\author{Rafael Rebolo}{address={Instituto de Astrof\'\i sica de Canarias, C/V\'\i a L\'actea, s/n, E38205 -- La Laguna (Tenerife), Spain}}

\begin{abstract}
Most of the known transiting extrasolar planets orbit slowly rotating
F, G or K stars. In here we report on the detection of a transiting
planet orbiting the bright, rapidly rotating A5 star HD15082
(WASP-33b, V=8.3, $m\,\sin\,i=86\,km\,s^{-1}$), recently made by
SuperWASP.  Time resolved spectroscopic observations taken during
transit show a hump caused by the planet crossing the line
profile. From the analysis of the spectra, we derive the radius of the
planet and find that it is orbiting retrograde in respect to the spin
of the star.  Because of its small distance from an A5 star (the
orbital period of only 1.22 days), the equilibrium temperature of the
planet is estimated to be 2700 K. The planet thus is one of the
hottest planets known, which makes it relatively easy to detect it in
the IR. We thus tried to detect it using the TNG but did not succeed.
\end{abstract}

\maketitle

%%%%%%%%%%%%%%%%%%%%%%%%%%%%%%%%%%%%%%%%%%%%
%% MAINMATTER
%%%%%%%%%%%%%%%%%%%%%%%%%%%%%%%%%%%%%%%%%%%%

\section{The importance of studying close-in planets of A-stars}

Studies of transiting extrasolar planets are of key importance for
understanding the nature of planets outside our Solar System, because
they allow to derive their mass, diameter and their density. The more
than 100 transiting planets discovered up to now thus give us a wealth
of information about the structure and evolution of extrasolar
planets. However, almost all of them are orbiting slowly rotating F, G
or K stars. Our knowledge about the planets of earlier type stars thus
is very limited.

It would however be very interesting to study planets of early type
stars, particularly transiting ones.  According to theory, the
frequency of massive planets depends on the mass of the star. The
frequency is expected to be higher for stars of higher mass than for
stars of lower mass.  Observations of planets of giant stars seem to
confirm this result (Johnson et al. 2010; Kennedy \& Kenyon
2008). However, it is not easy to determine the mass of a giant star,
let alone to estimate which mass it had when it was still on the main
sequence.  It would thus be important to confirm this results by
detecting planets of stars more massive than the sun but which are still
on the main sequence.

By comparing the properties of close-planets of A-stars with those of
late-type stars, we can learn a lot about the interaction between
stars and planets.  Close-in planets of G and K-type stars often have
radii that are too large for their mass. Such objects are usually
called ``inflated'' planets. It is generally believed that this kind
of inflation is caused by the interaction between stars and
planets. The discovery of CoRoT-9b supports this idea, since this
planet is {\em not} inflated, because it always has a relatively large
distance from its host star (Deeg et al. 2010).

Close-in planets are heated by tidal interaction and the radiation
from their host stars.  It has however, been shown that even if we take
these effects into account, it is still difficult to explain the large
sizes of the planets that are observed (Baraffe et al. 2003; Guillot
et al. 2010).  One possibility would be the presence of an additional
planet in an eccentric orbit. Such a planet would cause an eccentric
orbit of the inner planet, which would in turn lead to an
increased the amount of tidal heating. While this mechanism is very
attractive, it is unlikely that it works in all cases.

In order to better understand the interaction between stars and
planets, it would be ideal if we could simply exchange the central
stars with stars of different types, keeping all other parameters
constant. Since this is impractical, we observe systems containing
stars of different types instead. Since A-type stars are much hotter
than G-type stars, their planets will receive correspondingly more
photons in the visual regime. Whether this is also the case in the EUV
and the X-rays is not obvious. The amount of radiation which the
planet receives in this wavelength regime is important, because the
evaporation of the atmosphere of a planet depends on the amount of
radiation it receives shortward of the Ly-$\alpha$ line.  In the
classical view, A-stars do not have an outer convection zones and thus
do not have chromospheres, or a coronae. Although there is now growing
evidence that this classical view is not be strictly true, the
chromospheres and coronae of A-stars are certainly not like those of
later type stars (Simon \& Landsman 1997a; Simon et al. 2002; Hempel
et al. 2005; Schr{\"o}der \& Schmitt 2007).  The stellar wind of an
A-star is also different from that of G-type star.  It has been
suggested that A-stars have only a radiatively driven winds (Babel
1995).  Thus, if the unknown heating mechanism of close-in planets is
related to either the coronae, the stellar winds, or the magnetic fields
of the host stars, the amount of inflation would be different for
planets of an A- and G-stars.

In order to better understand the interaction between close-in planets
and their host stars, it is thus important to study close-in planets
of A-stars.  In here we report on the detection of a transiting planet
orbiting the bright, rapidly rotating A5 star HD15082 (WASP-33).

The discovery of HD15082b was presented by Cameron et
al. (2010). Because the detection and the properties of the planet
were already described in detail in this article, we will just briefly
summarize the results in here.

\section{A unique star/planet system}

The light curve of HD15082b (WASP-33b) obtained by SuperWASP shows a
flat-bottomed, planet-like transits recurring every 1.22 d (Christian
et al. 2006; Pollacco et al. 2006).  Further photometric monitoring
with the 0.95-m James Gregory Telescope (JGT) at the St.  Andrews
University Observatory, the 60-cm telescope of the University of
Keele, and 35-cm Schmidt-Cassegrain telescope at the University of
London Observatory at Mill Hill confirmed the transit, and allowed to
refine the transit-depth and the transit duration to a higher
accuracy.

Because we could not confirm the planet by radial-velocity
measurements, we used time resolved spectroscopic observations taken
during transit instead.  A transiting planet causes a bump that
crosses the line-profile during the transit. This bump distorts the
line-profile which results in a change of radial-velocity. This change
of radial-velocity during transit can even be measured for slowly
rotating stars (Rossiter-McLaughlin effect). If such a signature is
present and matches the transit geometry deduced from photometry, the
presence of a planet is confirmed.  Jenkins et al. (2010) have also
noted recently that the Rossiter-McLaughlin effect provides a powerful
method for distinguishing planets from blends in cases where classical
methods of confirmation cannot be used. For rapidly rotating stars,
the bump can be seen in the line-profiles directly. We obtained three
time-series of spectra during transit with echelle spectrographs.  One
series was taken with the 2-m-Alfred Jensch telescope at Tautenburg
observatory, another one with the 2.7-m Harlan J. Smith Telescope at
McDonald Observatory, and a third one with the
Nordic-Optical-Telescope (NOT) at the Observatorio del Roque de los
Muchachos.  We used LSD in order to compute an average line-profile
and modeled it with a transiting planet. By modeling the photometric
light-curve, and the spectroscopic times-series, we determined the
orbital parameters and the radius of the planet to a high
accuracy. From the analysis of the spectra, we find that the planet is
orbiting retrograde in respect to the spin of the star.

\begin{figure}[t]
\includegraphics[height=0.8\textwidth,angle=-90]{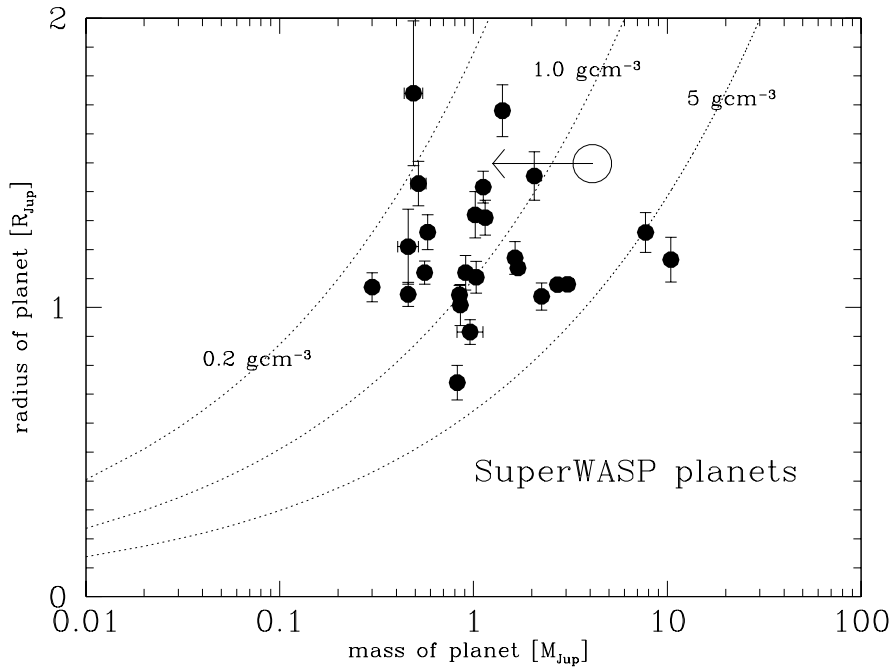}
\caption{Mass and density of the planets detected by SuperWASP.
HD15082b (WASP-33b) is marked as a circle.}
 \label{fig01}
 \end{figure}

We obtained 29 radial-velocity measurements with the Alfred Jensch
telescope in order to measure the mass of the planet.  Although an
iodine cell was used for the observations, the huge v\,sin\,i of the
star of 90\,$km\,s^{-1}$ limited the accuracy of our measurements to
about 700\,$m\,s^{-1}$.  Unfortunately, we could only derive an upper
limit for the mass of the planet of 4.1\,$M_{Jup}$. Combined with the
radius of $1.497\pm0.045\,R_{Jup}$, this gives us an upper limit for
the density of 1.2\,$ g\,cm^{-3}$. The planet thus may well belong to
the class of low-density planets, which would mean that the
evaporation rate could be high. Figure 1 shows the masses and radii
for the planets discovered by SuperWASP including HD15082b (marked as
a circle).

It thus turns out that the planet has unique properties (Table 1). The orbital
separation is only $0.02555\pm0.00017$ AU, or $3.83\pm0.09$
$R_{star}$.  Since the star has a temperature of $7430\pm100$ K, this
planet receives about five times as much radiation in the visual band
than a planet orbiting a solar-like star at the same distance.
Simon \& Landsman (1997b) studied the emission of many A to F stars at
1900\,\AA , and find that the relative flux of A-stars is on average
higher than that of late type stars. However, the flux at 1900\,\AA has
scatter of $\pm0.4$ mag for stars of the same spectral type. It is
thus not easy to estimate how much brighter the HD15082 is at
1900\,\AA \, than a solar-like star. It is possible that it is more
than 100 times brighter than a G-type star but we can not be certain
about that. Even more difficult is it to estimate the flux short-ward
of Ly-$\alpha$ ($\lambda=121.5668\,nm$) which is the essential
parameter in order to estimate the mass-loss of the planet.

% density ((4.1)*(1/(1.497**3)))*1.326 g cm-3=1.222+/-0.104 g cm-3
% distance modulus 5.32+/-0.28 mag

\begin{table}[t]
\begin{tabular}{ll}
\hline
\tablehead{1}{l}{b}{Parameter} & \tablehead{1}{l}{b}{Value} \\
\hline
Stellar mass 	                      & $1.495\pm0.031$ $Ms/ M_\odot$ \\
Stellar radius	                      & $1.444\pm0.034$ $Rs/ R_\odot$\\
Spectral Type	                      & kA5hA8mF4 \\
Distance 	                      & $116\pm16$ pc\\
Teff	                              & $7430\pm100$ K \\
log\,g	                              & $4.3\pm0.2$ \\
v\,sin\,i                  	      & $90\pm10$ $km\,s^{ −1}$ \\
$\rm [M/H]$	                      & $0.1\pm0.2$ \\
$M_B$                                 & $3.1\pm0.3$ \\ 
$M_V$                                 & $3.0\pm0.3$ \\ 
$M_J$                                 & $2.2\pm0.3$ \\ 
$M_H$                                 & $2.2\pm0.3$ \\ 
$M_K$                                 & $2.1\pm0.3$ \\ 
Epoch of mid-transit, Tc 	      & $2\,454\,163.223 73\pm0.000\,26$ HJD\\
Orbital period  	              & $1.219 8669\pm0.000 0012$ d\\
Orbital separation, a    	      & $0.02555\pm0.00017$ AU\\
Projected obliquity, $\lambda$        & $252\pm2^o$\\
Planet mass              	      & $<4.1$ $Mp/M_{Jup}$ \\
Planet radius            	      & $1.497\pm0.045$ $Rp/R_{Jup}$\\
\hline
\end{tabular}
\caption{Properties of HD15082 and HD15082b}
\label{table1}
\end{table}

\section{Spectroscopic observations of the secondary transit}

Because the planet is so close to a hot A-type star, at least the side
facing the star must be very hot. A simple model leads to a
temperature of about 2800 K but the real temperature could even be
higher.  This is very likely, since the true temperature of HD149026b
is also much higher than the calculated temperature using the same
simple model (Harrington et al. 2007).  If the temperature of HD15082b
is 2800 K, it should be possible to detect it.

During the secondary transit of an extrasolar planet, the star occults
the planet. Shortly before that moment, the planet is observed at its
maximum brightness. Thus, by observing the planet before, during and
after the secondary transit, it is possible to determine the amount of
radiation coming from the planet alone.  Such secondary eclipses have
been observed most notably from space (e.g. Charbonneau et al.  2005;
Deming et al. 2006; 2007) but have also been detected from the ground
(de Mooij \& Snellen 2009; Sing \& L\'opez-Morales 2009).

The other property that makes this object very special is that the
host star is unusually bright (J=7.58, H=7.52, K=7.47 mag).  This
opens up the possibility to observe directly the thermal emission of
this very hot Jupiter.

Up to now, the temperatures of the planets that have been studied are
typically about 1000 K (GJ436b: $712\pm36$K, Deming et al. 2007;
TrES-1: $1060\pm50$ K, Charbonneau et al. 2005; HD189733b: $1117\pm42$
K, Deming et al. 2006; HD209458b: $1130\pm150$ K, Deming et al. 2005;
HD189733: $973\pm33$ K, Knutson et al 2007). Up to now, there are only
two planets for which the evaporation exosphere has been detected:
HD209458b, and HD189733b. Observation of the Ly-$\alpha$ line of
HD209458b with the HST allowed to estimate the evaporation rate to be of
the order of $10^{10}$\,$g\,s^{-1}$ (Lecavelier des Etangs 2009).

In order to learn more about the properties, the evolution and
particularly the the evaporation of close-in extrasolar planets it is
best to study the most extreme cases. How hot are these planets, and
how long can they survive the intensive heat of their host stars? Is
it possible that hot Neptunes are the cores of evaporated Jupiters?
Observations of HD15082b could answer these questions.

Using 2800 K for the temperature for the planet and $1.5\pm 0.05$
$R_{\rm jup}$ for its radius, we find that the planet has to be
brighter than about $m_J=15.11$, $m_H=14.8$, $m_K=14.4$, or $\approx
$7 mag fainter than the star. The problem thus is not the brightness
of the planet itself but the brightness difference between the star
and the planet. Figure~2 shows the estimated brightness difference
between planet and star. For example, in the J-band the planet-to-star
flux ratio is about 0.001, corresponding to 1 mmag.  Figure 2 shows
the expected brightness ratio between the star and the planet. We
expect a that the flux of the planet is $\geq 0.1\%$ to $\geq 0.2\%$
in the NIR.

Using the Telescopio Nationale Galileo (TNG) at the Observatorio del
Roque de los Muchachos and its NICS spectrograph have have obtained
spectra in and out of transit.  We used the IJ-grism which covers the
wavelength range from 0.9 to 1.45\,$\mu m$. With a slit-width of 1
arcsec, NICS achieves a resolution of $\lambda
/ \Delta \lambda=500$. In order to minimize the flux-losses, we
opened the slit as far as possible, which is 2 arcsec. The object was
observed in-transit for about half an hour and out-of transit for one
hour. Unfortunately, the night was partly cloudy which limited the
time for which we could observe the object.

\begin{figure}[t]
\includegraphics[height=0.78\textwidth,angle=-90]{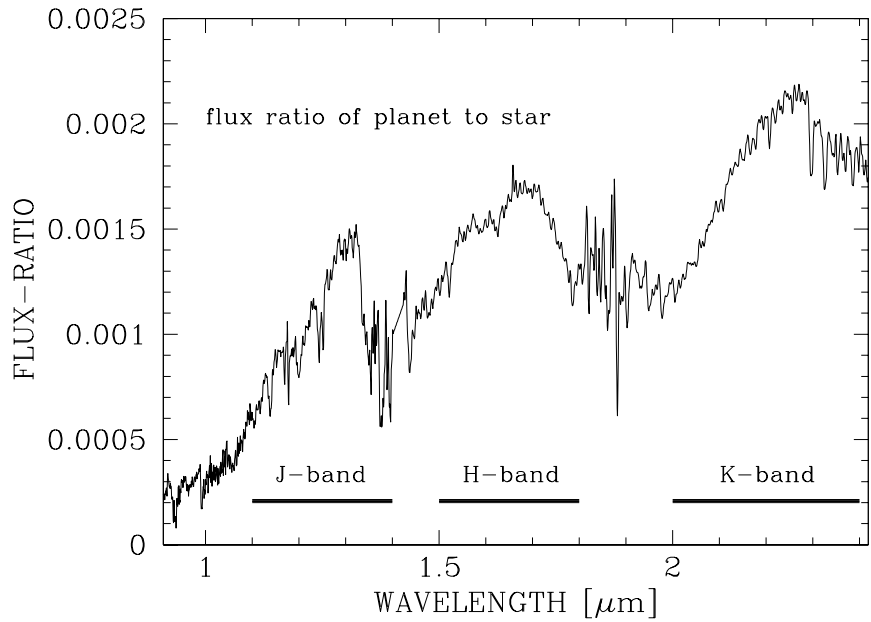}
\caption{Expected brightness-ratio between the star and the
planet.}
 \label{fig02}
 \end{figure}

\begin{figure}[h]
\includegraphics[height=0.78\textwidth,angle=-90]{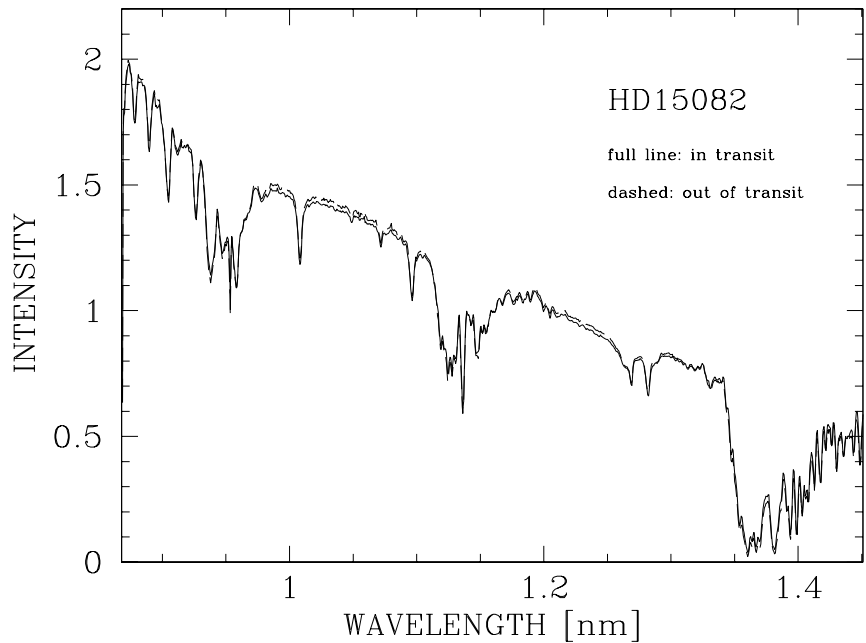}
\caption{Spectrum of HD15082b (WASP-33b) taken in and out of the
secondary transit. The secondary transit of the planet is not detected.}
 \label{fig03}
 \end{figure}

The spectra are shown in Figure 3. Unfortunately, due to clouds, we
can only derive an upper limit of 1.2\% for the relative flux of the
planet. This is clearly insufficient for the detection of the
planet. The upper limit for the temperature that we can derive is
roughly 5500 K. Spectra of higher signal-to-noise ratio or photometric
data of sufficient quality are needed in order to detect the secondary
eclipse of the planet.

\clearpage
%%%%%%%%%%%%%%%%%%%%%%%%%%%%%%%%%%%%%%%%%%%%%%%%
%% BACKMATTER
%%%%%%%%%%%%%%%%%%%%%%%%%%%%%%%%%%%%%%%%%%%%%%%%

 \begin{theacknowledgments}
We are grateful to the user support group of TNG for all their help
and assistance for preparing and carrying out the observations.
 \end{theacknowledgments}

\bibliographystyle{aipprocl} % if natbib is missing

% bibtex \jobname

\end{document}